# Transitioning STEM-focused Teacher Professional Development from f2f to Online

Carmen Fies
The University of Texas at San Antonio
USA
Carmen.fies@utsa.edu

Chris Packham
The University of Texas at San Antonio
USA and
National Astronomical Observatory of Japan
Japan
Chris.packham@utsa.edu

**Abstract:** This paper compares two cases of a Teacher Professional Development (TPD) focused on astronomy education: the San Antonio Teacher Training Astronomy Academy (SATTAA). The central question here is: How do in-service teachers' perceptions of the logistics and key benefits of SATTAA compare across two cases: the 2019 fully face-to-face (f2f) iteration in 2019, and the fully online iteration in 2020. Participants in both iterations equally indicated that they thought of their experiences as valuable and the program effective with two exceptions: (1) field trips that took place f2f were ranked higher than virtual options; and (2) technology was highlighted as benefit in the 2020 online iteration, but not in the 2019 f2f program.

## Introduction

COVID-19 caused a need to rapidly transition classroom practices and professional gatherings typically taking place in face-to-face (f2f) modalities to fully online experiences. The expediency of that shift pushed logistic limits of educational institutions. This paper reports on the transition of the San Antonio Teacher Training Astronomy Academy, or SATTAA (Fies & Packham, 2020; Fies et al., 2021; Packham et al., 2020), from a f2f to an online program. The central question addressed here is: How do in-service teachers' perceptions of the logistics and key benefits of SATTAA compare across two cases: the 2019 f2f iteration and the 2020 online iteration?

In order to better situate the discussion that follows, we provide an overview of SATTAA here. The theory of change that informs our program rests on two core insights. First, that children want to understand the world they live in and are innately curious about many of the things scientists study, such as the night sky, rockets, or dinosaurs. Yet, somewhere during their teens, many students become disenchanted with learning about scientific ideas and processes. Secondly, teachers serve in crucial roles as they work with students. Not only are they formal educators, but also mentors and guides who help shape students' thinking beyond the knowledge of a scientific discipline they teach. Teachers thus are uniquely well suited to foster a sense of enthusiasm about science in their students. Therefore, high-quality Teacher Professional Development (TPD) directly connects with a major point of leverage in contributing positively to STEM learning.

SATTAA seeks to provide teachers with opportunities for the acquisition of additional content expertise towards more confident teaching of astronomy, while also providing pedagogical and technological tools and resources to make the teaching of those classes fun and interesting. In SATTAA, astronomers engaged in astrophysics research and higher education design the content sequence. However, intertwining this content with productive pedagogies and technologies in secondary classrooms is of equal importance, and accomplished through team members with expertise in teacher preparation in general, and STEM education in particular.

## Theoretical Foundation

Our approach to SATTAA is grounded in sociocultural constructivism (e.g., Vygotsky, 1962, 1978). Specifically, the design rests on the notion that learning takes place by building on prior knowledge, is extended through action and interactions with others and with the content at hand, and is situated in and influenced by the social context in which the learning takes place. The design is further informed by the TPACK Model (Mishra & Koehler, 2006), as participants' experiences intertwine the building of content knowledge with building pedagogical and technological expertise.

## Literature

Since SATTAA is centered on astronomy education, we offer a brief statement of why it is necessary to study all facets of astronomy education research (AER). The existing literature shows promising directions, but also substantive gaps. For example, Bretones (2019) notes the critical need for AER, and advocates for a broad approach that includes all stakeholders, specifically teachers. Future research must include the lenses that help to better understand the teaching and learning of astronomy content from early childhood through terminal degrees in the field. This includes the study of TPD, which is of particular interest as the majority of astronomy teaching in Texas occurs as an out-of-field teaching assignment, a practice that Ingersoll (1999) noted as prevalent and detrimental.

### Teacher Professional Development

Avalos' (2011) review of TPD research identifies four recurrent themes: general professional learning, mediations, conditions and factors, and effectiveness studies. The author concludes that the literature evidences "that teacher learning and development is a complex process that brings together a host of different elements and is marked by an equally important set of factors" (p. 17). Avalos notes that, although it is unclear how persistent TPD-related gains are, "prolonged interventions are more effective than shorter ones, and … combinations of tools for learning and reflective experiences serve the purpose in a better way" (p. 17). Darling-Hammond (2017) points out policy considerations that surround education: "Because of inequalities in US school funding, teacher salaries and ongoing supports, those who teach in poorer districts are also less likely to receive ongoing professional development, which further exacerbates inequality in students' access to quality teaching" (p. 293). This plays a critical role in any setting where access to TPD is inequitable based on school district funding or policy.

A model sensitive to the productive inclusion of digital technologies is the Technological Pedagogical Content Knowledge, or TPACK, framework (Mishra & Koehler, 2006) which posits that teachers' competencies must include three facets, individually and in their interactions with each other: content, pedagogical and technological knowledge. Studies of TPACK span a wide variety of content areas and educational settings (e.g., Agustini, 2019; Borthwick & Pierson, 2008; Kim, Kim, Lee, Spector, & DeMeester, 2013; Lee & Kim, 2014; MaKinster, Trautmann, & Barnett, 2014; Murthy, Iyer, & Warriem, 2015; Sun, Strobel, & Newby, 2016; Zwiep & Benken, 2012). The technology focus here is on integration, and includes general technology skills.

### Transitioning to Online Practice

There is a great need for publications that thoroughly capture the rapid-speed adjustments necessitated by the pandemic. However, there is prior research investigating TPD specifically with a lens of shifting to online learning. Noting that synchronous online professional learning is considered by teachers as valuable and with the potential to also foster a sense community, McConnell et al (2013) describe such TPD as a viable option as long as technological challenges are addressed in the planning phase. Canuel and White (2014), in a study of transitioning to a virtual high school, conclude that, not only is it necessary to offer ongoing TPD rather than brief workshops, but "there is an urgent need for effective professional development for educators in virtual schools" (p. 178). In a Saudi Arabian study of TPD focused on science teachers (Binmohsen & Abrahams, 2020), results indicate that a fully online program can not only achieve equivalent results as a f2f iteration, but may offer additional benefits. As COVID-19 externally motivated a need for adaptation, the driving question became how to adapt most effectively to meet goals and objectives. This, in part, addressed a finding in the literature tied to teachers' motivations, and hesitations, to shift their practice toward inclusion of a new or different technology tool in their classroom practice.

For example, Kim et al. (2013) suggest that "to help teachers understand that beliefs other than those they have acted on in the past … might be more effective in achieving something that they already want to achieve" (p. 84).

## Methodology

We compare two iterations of the same astronomy education focused TPD program: SATTAA. In 2019, the 2-week program took place f2f; in 2020, it took place completely online. As each case is bounded by the modality and the year of the TPD, the unit of analysis is the iteration. The methodological home of this paper is situated in case study research (Compton-Lilly, 2012; Merriam, 1988), with the goal to describe the essential features of logistics and key benefits, as perceived by participating in-service teachers.

### Context

SATTAA's home is San Antonio, a large urban center in Texas with roughly 1.5M inhabitants and a minority-majority population (United States Census Bureau, 2020). While 'white alone, not Hispanic or Latino' accounts for 24.8% of the overall population, with 64.2%, the largest demographic is Hispanic or Latino. 18.6% of the population are listed as being in poverty. Out-of-field teaching of astronomy is the rule rather than exception in Texas schools. Drawing on data collected by the Texas Education Agency (Smith, 2020), only 5.22-6.23% of all new science teacher certifications annually issued between 2016-19 were specific to any form of physics content, which we use as proxy to astronomy content expertise. SATTAA, a formal education, informal education, and research institution partnership, aims to provide teachers opportunities to strengthen their astronomy content knowledge integrated with explicit connections to pedagogy, technologies, and the state standards for schools. In 2019 and 2020, SATTAA served pre- and in-service teachers. In addition to content sessions, teachers participated in field trips, integrated what they learned in the design of future astronomy lessons for their own classrooms, and completed NASA Universe of Learning tasks.

### Participants: In-service Teachers

The participating in-service teachers were predominantly female (60% in 2019, and 73% in 2020). In each year, all of the participating teachers worked at different area schools, and, with the exception of two 2020 teachers who taught at parochial schools, all worked in public schools. In addition, the participants in both iterations worked at schools with a higher proportion of underrepresented minority students and a higher proportion of students are identified as having a greater economic need than is the case in San Antonio's overall population (Tab. 1).

| Year | Teachers, female (n) | Teachers, male (n) | Students: % White | Students: % Economically Disadvantaged | Students: % English Learners | Students: % Special Education |
|------|------|------|------|------|------|------|
| 2019 | 3 | 2 | 15.3 | 55.9 | 10.0 | 11.3 |
| 2020 | 11 | 4 | 11.4 | 62.9 | 9.8 | 10.0 |

*Table 1: In-service teachers' school demographics (percent averages).*

Whereas 24.8% of the city-wide population is identified as white, the proportion of white students in the participants' schools is considerably smaller (15.3% in 2019, and 11.4% in 2020). In addition, students at these schools are identified as in need at a higher rate than is the city-wide poverty rate of 18.6%. Across the 2019 participants' schools, 55.9% of students identified as economically disadvantaged. The 2020 participants' schools showed an even higher rating at 62.9% (Fig. 1).

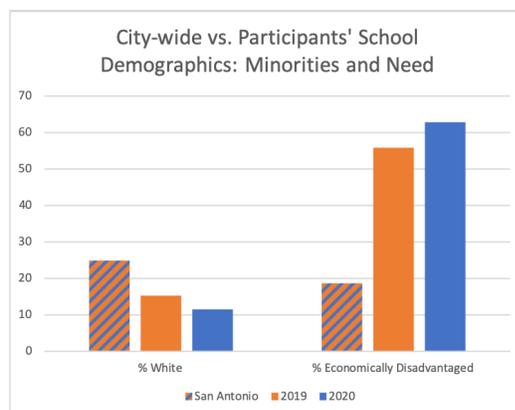

*Figure 1: Comparison of participants' school demographics with city-wide demographics. Students at these schools are of a higher proportion members of minority populations, and are more likely to be in economic need than is the case city-wide.*

## Results

### Logistics

We categorized the schedules of both iterations by session type and role in the overall TPD to develop a matrix of all event elements, identifying what remained unchanged and what was adapted. This served as basis for a *percent-time comparison* between the two years.

*SATTAA2019 (f2f)*: For the majority of experiences, participants attended 4-hour morning sessions at the university campus and afternoon sessions either were facilitated online by an out-of-state instructor, or consisted of field trips. Two of the three field trips were local, and one a 6-hour drive away. For local field trips, participants met on site; the field trip to the McDonald Observatory consisted of a 3-day experience. Materials were either handed out as hard-copies, or on a thumb drive, and also were available via Google Drive. The contact-hour time distribution of the TPD was heaviest weighted toward content learning experiences (63%). Predominantly, the remainder of the time was split between logistics (21%), which includes driving time for the most distant field trip, and pedagogy-focused sessions and teachers' presentations at the end of the two weeks (16% of the overall time).

*SATTAA2020 (online):* Initially, the program had been planned as a f2f session, largely mirroring the 2019 iteration; however, once it became apparent that f2f was no longer an option, the redesign resulted in a fully online experience taking place via Zoom video conferencing. To make joining sessions as easy as possible, all participants and facilitators received a full program and calendar invites for all zoom sessions ahead of time. In addition, daily follow-up emails included the zoom link for the next morning meeting. Participants could access all files digitally during the event, and still have access to the shared repository on the Google Drive, and to session recordings hosted on a YouTube channel with protected access rights. In 2020, prior content topics could be retained and somewhat extended; even field trips could be retained to a degree. Through collaborative redesign conversations, the two already planned local field trips were adapted to online interactions. Unfortunately, there was no option to do the same with the field trip to the McDonald Observatory, a highlight of the two prior years; another virtual local field trip was added instead. To support facilitators, zoom practice sessions were offered (e.g., explorations of shared whiteboard use, breakout rooms, and polling), and sessions were moderated. The contact-hour distribution again was heaviest weighted toward providing content learning experiences and, at 72%, to an even greater extent than in 2019. Since the time component for logistics during the TPD was reduced to 4%, more time was also dedicated to pedagogy-focused engagement and the teachers' final presentations (24%).

### Perceptions

The participants' oral and written contributions were coded and analyzed in NVivo, following the constant comparative method outlined by Strauss and Corbin (1998) as it represents a systematic process of letting themes emerge from participants' utterances through cycles of open, axial, and selective coding.

*Logistics, including technology.* In 2019, the teachers indicated satisfaction with the schedule, with a single suggestion for change: to replace afternoon online session with f2f interactions. This indicates that the cohort felt f2f interactions were more beneficial to them than online sessions in the program. All teachers stated that the program's content should not be changed. A highlight was the field trip to the observatory; even though the drive was long and the schedule ambitious, all participants felt particularly enriched by the experience. The only difficulties noted were due to in-session technology-use, and mostly specific to a task that was perceived to have a steep learning curve. In 2020, the overall feedback was enthusiastic; again, teachers stated that TPD timing and duration were appropriate. Several noted that if SATTAA had not been offered fully online, they could not have participated regardless of COVID-19. All participants noted that session structure, feedback cycles, and virtual field trips were beneficial, and that all of these elements should be retained for future program iterations. However, overall ratings of virtual field trips were slightly lower than those for f2f field trips in the year prior. Technology was noted as challenge, including as necessary for specific tasks and as backbone of communication and shared resources.

*Key benefits*: All but one of the 2019 teachers indicated that the TPD was 'very relevant' as related to their future teaching; the one dissenting voice ranked it as 'relevant'. Coding of the teachers' mentions of benefits indicate that the highest impact was in terms of *resources* and *pedagogical* tools that could be seamlessly integrated in existing practice (33.3% and 26.7% of statements, respectively), followed by *community building and networking* (20%). *Field trips* were highlighted in 13.3% of the statements, and the remaining 6.7% were specific to *content*. None of the utterances noted *technology* as key benefit. 2020 participants unanimously rated the TPD as 'very relevant' for their future teaching, and indicated that they planned to share what they learned with colleagues. As was the case in 2019, the highest key benefit coding frequency was found to be for *resources* and *pedagogical* tools (26.6% and 20.9%, respectively). However, *content* knowledge ranked third with 16.5% of coding instances, closely followed by *community building and networking* with 15.8%. *Technology* was mentioned as a key benefit in 13.7% of the instances, and *field trips* in only 6.5%.

*Comparing the two cases through the lens of in-service teachers' perceptions:* The participants rated both iterations as highly useful for their future teaching of astronomy, and identified opportunities to collaborate with others as a critical component of the TPD. The teachers expressed a desire for more SATTAA opportunities, and felt not only better prepared to teach the content, but also more excited and passionate about it. Not entirely unexpected, virtual field trips were not rated as highly as f2f opportunities. However, the most marked difference in the online iteration was that teachers highlighted the benefit of technology use in the TPD on three levels: as a tool of *content* learning, as a tool of *communication* within the TPD, and as tool to support their *own learners in online settings*. That this aspect was not foregrounded in feedback to the 2019 iteration is likely a by-product of the standard classroom practice prior to COVID-19. In 2020, teachers had just finished teaching online and were more motivated to explore how technology may be able to support learning.

Given that our data indicate that our online iteration was at least as effective as the f2f iteration, we plan to keep SATTAA a virtual TPD with the hope that it will make it possible for teachers who are a greater distance from San Antonio, for example those working at rural schools, to equally benefit from the program. We are particularly encouraged that the teachers who participated in both years presented here work at schools that serve student populations with higher minority and economic need demographics than is the case city-wide. That the trend strengthened in the fully online year tells us that we are not only able to maintain, but to even expand the program's reach in a virtual setting. We note the importance of diversity and inclusion to both NASA's and NSF's mission, which aligns exceptionally well to the goals, vision, and results of SATTAA. Last but not least, we also see as an important byproduct of the online shift that the event creates a smaller carbon footprint. Although the difference between f2f and online iterations in this program is much smaller than it would be for a large conference, it is a contribution to reducing the human impact on climate change (Burtscher et al., 2020).

## Acknowledgements.

We thank NASA, the Space Telescope Science Institute, and NSF grant no. 1616828 for supporting this work financially. We also gratefully acknowledge the generous support of BEAT LLC (https://beatllc.com) for providing the assistance to enable SATTAA grants starting 2020. We also thank the many colleagues who have contributed to the program in each year.